\documentclass[twocolumn,showpacs]{revtex4}

\usepackage[latin1]{inputenc}
\usepackage{graphicx}%
\usepackage{dcolumn}
\usepackage{amsmath}

\makeatletter
\def\btt#1{\texttt{\@backslashchar#1}}%
\DeclareRobustCommand\bblash{\btt{\@backslashchar}}%
\makeatother

\begin{document}


\title{Microscopic  features of moving traffic jams}
\mark{Microscopic  features of moving traffic jams}

\author{Boris S. Kerner $^1$,  Sergey L. Klenov $^2$,  Andreas Hiller $^3$, and Hubert Rehborn $^4$}


\affiliation{$^1$, $^3$, $^4$
DaimlerChrysler AG, REI/VF, HPC:  G021, 71059 Sindelfingen, Germany 
}

\affiliation{$^2$
Moscow Institute of Physics and Technology, Department of Physics, 141700 Dolgoprudny,
Moscow Region, Russia
}



\begin{abstract}
 Empirical and numerical  microscopic features of moving traffic jams
are presented. Based on a single vehicle data analysis,
it is found that within wide moving jams, i.e., between the  upstream and downstream jam fronts
 there is a complex microscopic spatiotemporal structure. This jam structure consists of alternations of regions in which
 traffic flow is interrupted and flow states of low speeds associated with $\lq\lq$moving blanks" 
 within the jam. Empirical features of the moving blanks are found.
 Based on microscopic models
 in the context of  three-phase traffic theory, physical reasons for moving blanks
 emergence within wide moving jams are disclosed.
  Structure of moving jam fronts is studied based in microscopic traffic simulations.
Non-linear effects
associated with moving jam propagation 
 are numerically investigated and compared with empirical results. 
  \end{abstract}

\maketitle

\section{Introduction}
\label{Intr}

Freeway traffic is a complex dynamic spatiotemporal process.
A huge number of traffic flow models have been introduced for explanations of traffic phenomena  
(see the book~\cite{KernerBook}, the reviews~\cite{Gartner,Wolf,Sch,Helbing2001,Nagatani_R,Nagel2003A,MahnkeRev},  and the conference proceedings~\cite{Mah,SW3,SW4,SW5}).

In empirical freeway traffic observations, it has been found that traffic can be either free or congested.
In congested traffic, the $\lq\lq$stop-and-go" phenomenon is observed, i.e.,
a sequence of moving jams can appear
(see the classic papers by Edie and Foote, 
Treiterer et al., Koshi et al.~\cite{Edie1958,Edie1960A,Edie1961A,Treiterer1974A,Treiterer1975,Koshi}).
A moving jam is a localized structure propagating upstream. The jam is spatially restricted by two jam fronts in which
the speed, density, and flow rate vary sharply.
Within the upstream jam front, vehicles must decelerate to the  speed in the jam.
Within the downstream jam front, 
vehicles 
accelerate escaping from the jam.
 
Recently, from spatiotemporal analysis of 
empirical data measured over many days and years on various freeways in different countries, Kerner found that
there are two different phases in congested traffic, synchronized flow and wide moving jam
(see references in the book~\cite{KernerBook}). Thus, 
there are three traffic phases: 1. Free flow. 2. Synchronized flow. 3. Wide moving jam. 

The fundamental difference between
synchronized flow and wide moving jam is determined by the following macroscopic spatiotemporal objective (empirical)   
  criteria, which define the phases~\cite{KernerBook}. 
The wide moving jam is a moving jam that exhibits the
\emph{characteristic, i.e., unique, and coherent feature}  to maintain the mean velocity of the downstream jam front, 
 even  when the jam propagates through any other
traffic states or  freeway bottlenecks.   
In contrast, 
 synchronized flow  does not exhibit this characteristic feature, in particular, the downstream front of synchronized flow
is often  \emph{fixed} at a freeway bottleneck.

  Wide moving jam propagation through a freeway bottleneck is associated with traffic flow interruption within the 
  jam: Vehicles
  are in a stop within the jam during a time interval, which is considerably longer than the mean  time delay in vehicle acceleration at the
  downstream jam front. Thus,   flow interruption within  wide moving jams 
  can be used as a microscopic criterion for distinguishing between
  the wide moving jams and synchronized flow in congested traffic, even if single vehicle data is measured at a single freeway 
  location~\cite{KKH}. The traffic flow interruption effect discloses the  physical nature of a qualitatively
different behavior of the wide moving jam phase  in comparison with the synchronized flow
phase. 
    Moreover, from an analysis of wide moving jams it can be assumed that there are moving blanks within the jams:
  Vehicles come to a stop at different net distances (space gaps) to each other at the upstream front of a wide moving jam. 
  Later, vehicles
  begin to move within the jam to cover these blanks within the jam; as a result blanks appear that move  against the flow. 
  
 It can be expected that both
   flow interruption within the jams 
  and moving blanks should influence  microscopic features and characteristics of moving jams considerably.  
  However, a  microscopic spatiotemporal structure
  of moving jams, in particular, the   effect of moving blanks as well as
  microscopic features of the jam fronts have   not been  understood. 
In this paper, based on an empirical single vehicle data analysis the microscopic spatiotemporal structure
  of moving jams and its microscopic features have been found. These empirical features have been explained
  using numerical simulations of a microscopic model in the context of three-phase traffic theory, which
   is adequate with all known empirical observations of congested traffic~\cite{KernerBook}.
 The article is organized as follows. 
Empirical and theoretical features of moving blanks within wide moving jams   are  considered in Sect.~\ref{Blanks_Sect}.   
  In Sect.~\ref{Front_Sect}, microscopic features of the jam fronts and jam propagation are numerically studied. 
  
  \section{Moving Blanks within Wide Moving Jams}
\label{Blanks_Sect}

\subsection{Microscopic Characteristics of Moving Blanks in Empirical Single Vehicle Data}
\label{Em_B_Sect}

\subsubsection{Congested States in Single Vehicle Data}
\label{Con_states}

Single vehicle characteristics are usually obtained either in driver experiments or through the use of 
detectors (e.g.,~\cite{Neubert,Knospe2002,Knospe2004,Cowan,Koshi,Luttinen,Bovy,Tilch2000,Banks,Gurusinghe2003A}).
In the latter case, data from a single freeway location or aggregated data measured at different locations 
are used~\cite{Aggregated}.

\begin{figure*}
\begin{center}
\includegraphics[width=12 cm]{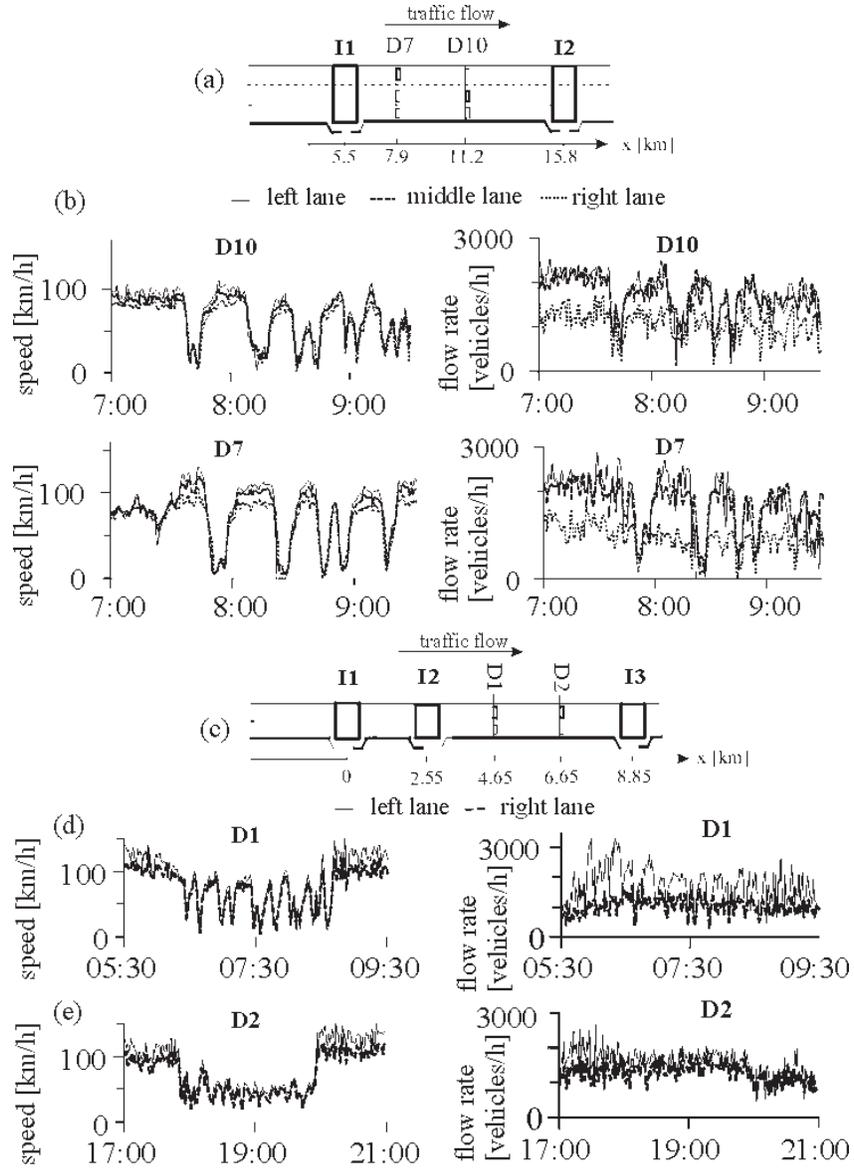}
\caption{Macroscopic characteristics of empirical single vehicle data measured on the freeway A5-South
(a, b) and on the freeway A92-West (c--e): (a) -- Sketch of detector arrangement of a section on the freeway A5-South.
(b) -- Local traffic dynamics (one-minute average data) of the speed (left) and flow rate (right)  
in three freeway lanes on December 12, 1995. (c) -- Sketch of detector arrangement on a section of the freeway A92-West.
(d, e) -- Local traffic dynamics (one-minute average data) of the speed (left) and flow rate (right) on   July 17, 2000 (d)
and August 23, 2000 (e)
in both freeway lanes.
 \label{Single_Fig} } 
\end{center}
\end{figure*}

Single vehicle data used in the article have been measured on two different freeways in Germany.
The first single vehicle data  set has been measured on the three-lane freeway section of the freeway A5-South  at detectors 10 and 7
(Fig.~\ref{Single_Fig} (a)); a detailed consideration of this freeway section is made in~\cite{Kerner2002B}.
For an overview of moving jams observed in single vehicle data and used below, the  data averaged over 1-min intervals are shown
in Fig.~\ref{Single_Fig} (b). The second set of single vehicle data has been measured
 on a two-lane section of the freeway A92-West at detectors  D1 and D2 
   between intersections $\lq\lq$AS Freising-S\"{u}d" (I2)
and  $\lq\lq$AK Neufahrn" (I3)
near  Munich Airport   (intersection I1: $\lq\lq$AS Flughafen" in Fig.~\ref{Single_Fig} (c)).
 Local dynamics of the average speed and flow rate (1-min averaged data) for two typical
days at which congested traffic have been observed are shown in Fig.~\ref{Single_Fig} (d, e). 

In   both freeway sections,
 single vehicle data are measured through the use of two sets of double induction loop detectors. 
A    detector set consists of two detectors for each of the freeway lanes.
A detector registers a vehicle by producing  a   current electric pulse whose duration $\Delta t_{i}$
is related to the time taken by the vehicle to traverse the induction loop. This enables us to calculate 
the gross time gap between two vehicles $i$ and $i+1$ that have passed the loop one after the other
 $\tau^{\rm (gross)}_{i,i+1}$. There are two detector loops in each detector.
The distance between these loops is constant. This enables us to calculate the individual vehicle speed $v_{i}$
and estimate
the vehicle length $d_{i}=v_{i}\Delta t_{i}$ as well as the net time gap (time headway) between the vehicles $i$ 
 and $i+1$: $\tau_{i,i+1}=\tau^{\rm (gross)}_{i,i+1}-\Delta t_{i}$.

\subsubsection{Alternations of Flow Interruption and Moving Blanks within Moving Jams}
\label{Interruption}

\begin{figure*}
\begin{center}
\includegraphics[width=12 cm]{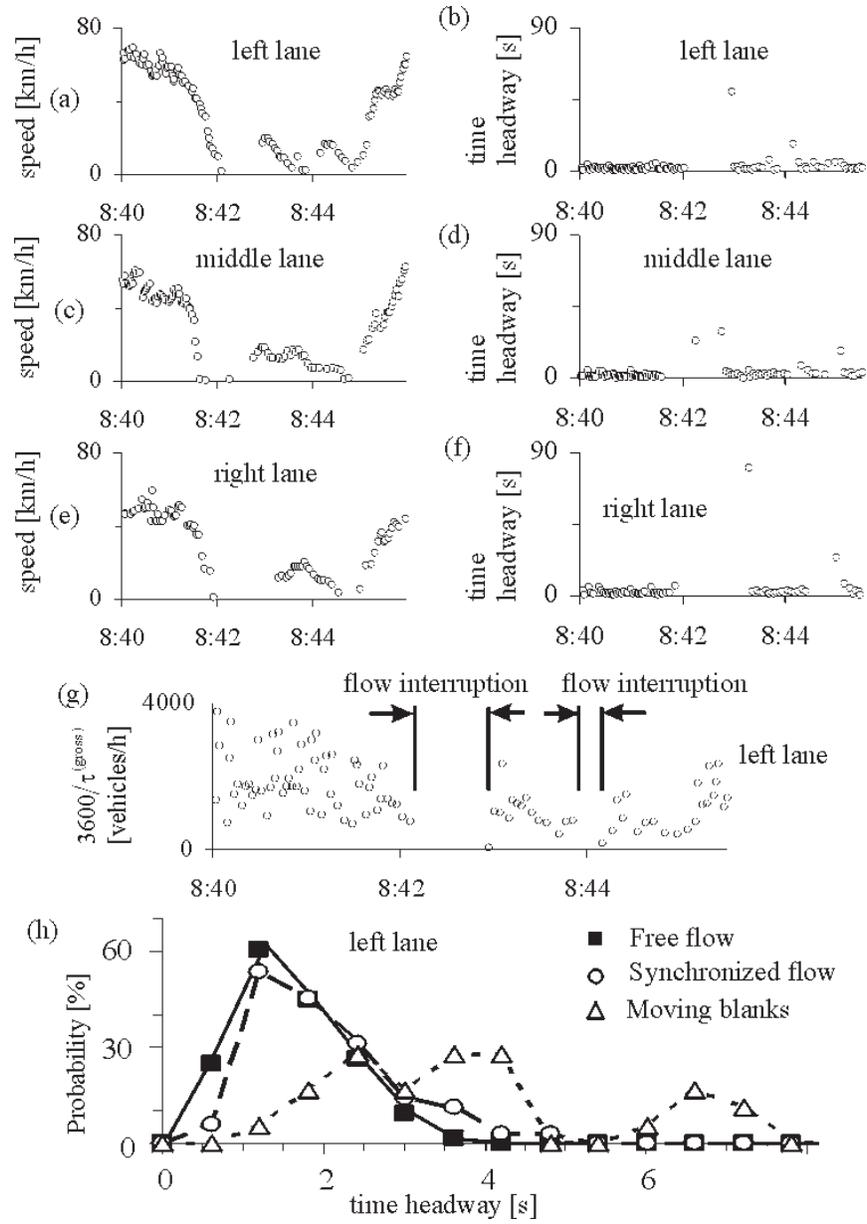}
\caption{Microscopic structure of wide moving jams:
(a--f) -- Empirical single vehicle data for speed within a wide moving jam  (a, c, e) and the associated
time headways (b, d, f)
in the left (a, b), middle (c, d),   and
right lanes (e, f) related to the example at D10 in  Fig.~\ref{Single_Fig} (b). (g) --   
  Time distributions of     the value $3600/\tau^{\rm (gross)}$ in the left lane. (h) --
  Probability for time headways in the left lane for
    free flow  (solid curve, $6:20\leq t\leq 6:40$),    synchronized flow
(dashed curve, $8:39\leq t\leq 8:42$);   moving blanks within the wide moving jam  (dotted curve, $8:42< t\leq 8:44$).
 \label{WideJam} } 
\end{center}
\end{figure*}

Within wide moving jams, i.e., between the jam fronts regions in which flow is interrupted
are alternated with moving blanks
   (Fig.~\ref{WideJam}). The condition for  flow interruption 
   is~\cite{KKH}:
\begin{equation}
\tau_{\rm max}\gg \tau^{\rm (ac)}_{\rm del},
\label{GrossDel}
\end{equation}
where $\tau_{\rm max}$ is the maximum time headway
    between  two vehicles within the   jam and $\tau^{\rm (ac)}_{\rm del}$ is the mean time
delay   in vehicle acceleration  at the downstream jam front  
from a standstill 
state within the jam. 
Under the condition (\ref{GrossDel}), there are at least several vehicles   within the jam that    are
in a standstill or if they are still moving, it is only with a negligible low speed in comparison with the speed
in the jam inflow and outflow. These vehicles separate vehicles accelerating at the downstream jam front from
vehicles  decelerating at the upstream jam front: The inflow into the jam has no influence on the jam outflow. Then
the jam outflow is fully determined by vehicles accelerating  at the downstream jam front.
In the example,
   flow interruption effect occurs two times during  jam propagation through
   detector D10~\cite{MovingJams} (Fig.~\ref{WideJam} (a--g);
   these time intervals are labeled $\lq\lq$flow interruption" in Fig.~\ref{WideJam} (g)). 
The   values $\tau_{\rm max}$ for the first flow interruption intervals within the wide moving jam
  are equal to approximately 50 s in the left lane, 30 s in the middle line, and 80 s in the right line.
  These values $\tau_{\rm max}$ satisfy the criterion (\ref{GrossDel}) because 
 corresponding to empirical results $\tau^{\rm (ac)}_{\rm del}\approx 1.5-2$ sec~\cite{KernerBook}.
   
Whereas the first flow interruption interval
  appears almost simultaneously in all freeway lanes,
the second one occurs in the left lane earlier than in the middle and right lanes 
   (Fig.~\ref{WideJam} (a--g)). In other empirical examples of wide moving jams,
 the flow interruption effect occurs in
  some of the freeway lanes only whereas in other lane(s) there is no flow interruption. It turns out that even in this case
 a moving jam can be a wide moving one, i.e., the jam propagates through
 a bottleneck while maintaining the mean velocity of the downstream jam front,
 if
 the relation $\tau_{\rm max}/\tau^{\rm (ac)}_{\rm del}$ for the lane(s) with flow interruption
 within the jam is great enough~\cite{KKH}. Similar results are found for other moving jams in single vehicle data, i.e., for the example   
  measured on the freeway A92-West (Fig.~\ref{Single_Fig} (d)).

\begin{figure*}
\begin{center}
\includegraphics[width=12 cm]{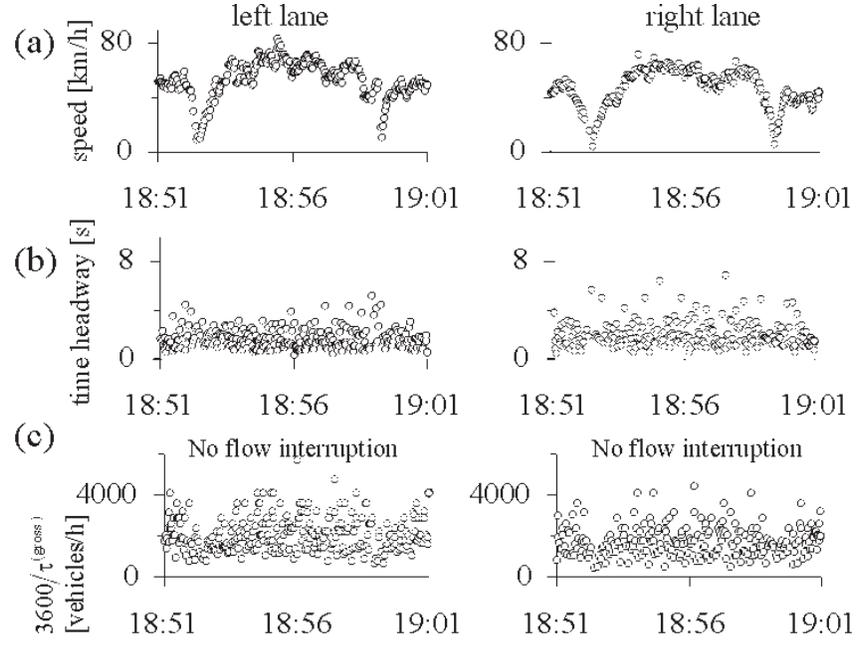}
\caption{Microscopic structure of narrow moving jams:  
Empirical single vehicle data for speed within a sequence of two narrow moving jams    
 in the left (left)   and
right lanes (right) (a) as well as
  the associated time distributions of time headways $\tau$ (b)
  and  of the value $3600/\tau^{\rm (gross)}$ (c)
  related to   Fig.~\ref{Single_Fig} (e).
 \label{NarrowJam} } 
\end{center}
\end{figure*}

 In other example  shown in  Fig.~\ref{Single_Fig} (e), there are also many moving jams during the time intervals of congested traffic.
 The speed  within  two moving jams in Fig.~\ref{NarrowJam} 
 is also very low.
 Nevertheless,
 rather than wide moving jams  these moving jams should be classified as
narrow moving jams~\cite{KernerBook,KKH}. This is because there are no traffic flow interruptions within  
these moving jams (Fig.~\ref{NarrowJam}). Indeed,
upstream and downstream of the jams, as well as within the jams there are many vehicles that traverse the induction loop detector:
There is no qualitative difference in the time-dependences of  time headways for different
time intervals associated with these narrow jams and in traffic flow upstream or downstream of the  jams (Fig.~\ref{NarrowJam} (b, c)).
This can be explained if it is assumed that each vehicle, which meets  a narrow moving  jam,  must
decelerate at the upstream jam front down to a very low speed within the jam, 
 can nevertheless accelerate at the downstream jam front   almost without any time delay within the jam: The narrow moving jam consists of 
 upstream and downstream jam fronts only.   These assumptions are confirmed by single vehicle data shown in Fig.~\ref{NarrowJam}, in which
 time intervals between different measurements of  time headways and for the value $3600/\tau^{\rm (gross)}$ 
  for different vehicles exhibit the same behavior
 away and within the jams~\cite{KKH}.
Thus, regardless of these narrow moving jams traffic flow is not discontinuous, i.e., the narrow moving jams belong indeed to the 
synchronized flow  phase.

  A comparison of  the microscopic criterion  
   (\ref{GrossDel})   with the macroscopic spatiotemporal criteria for the phases
made in~\cite{KKH} allows us to suggest that
the  moving jam in Fig.~\ref{WideJam} for which the criterion (\ref{GrossDel}) is satisfied
 is associated with the   wide moving jam phase: The jam propagates through a bottleneck
 while maintaining the mean velocity of
 the downstream jam front. In contrast, narrow 
moving jams in Fig.~\ref{NarrowJam}  for which the criterion (\ref{GrossDel}) is not
 satisfied are   associated with   the synchronized flow  phase: The narrow moving jam is caught at the bottleneck.

  Between flow interruption intervals within the wide moving jam in Fig.~\ref{WideJam},
   vehicles within the jam exhibit   time headways 
  about 5 sec or longer.
 The latter  can be explained by moving blanks within the jam.
 Empirical single vehicle data  (Fig.~\ref{WideJam}) shows that when vehicles
meet the wide moving jam, firstly they decelerate at the  upstream jam front sharply 
up to a standstill. As a result,
the first flow interruption interval in all lanes appears. It can be assumed that     
net distances  (space gaps) between these vehicles can be very different and
the mean space gap  can exceed  a minimum (safe)
     space gap considerably. For this reason, later the vehicles become to decrease
     these space gaps. As a result, low vehicle speed states appear within the jam
     as this can be seen in Fig.~\ref{WideJam} (a, c, e). Consequently, due to this vehicle motion 
     new space gaps, i.e., blanks between vehicles occur upstream within the jam. 
     Then other vehicles that are upstream  begin to move
     within the jam covering these blanks. This lead to occurrence of
     moving blanks   propagating upstream within the jam. 
     
     These assumptions
     are confirmed by low speed states  within the jam
     (Fig.~\ref{WideJam} (a, c, e)). These low speed states within wide moving jams
     associated with  moving blanks   are not  necessarily synchronized between different lanes.
     For example, in Fig.~\ref{WideJam} (a, c, e), in the interval 8:43--8:44
         there can be seen non-interrupted
     traffic flows of  low speeds in 
     the middle and right lanes associated with moving blanks within the jam,
      whereas in the left lane traffic flow is interrupted.
      Mean time headways related to moving blanks within the jam are usually longer (3--7 sec) than
      they are away from the jam in free flow and in synchronized flow (Fig.~\ref{WideJam} (h)).

\subsubsection{Empirical characteristics of traffic phases}

\begin{figure*}
\begin{center}
\includegraphics[width=12 cm]{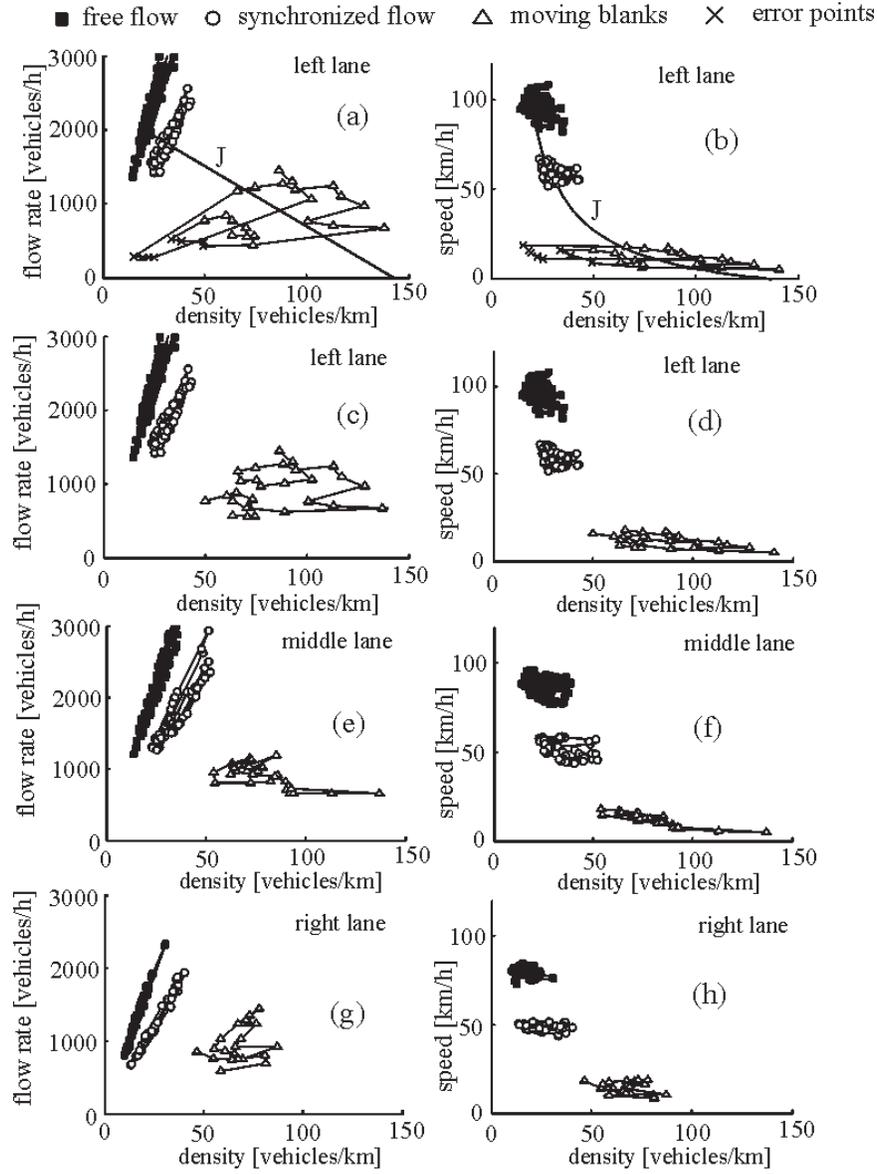}
\caption{Empirical characteristics of traffic phases: (a, b) --
Traffic phases in the flow--density (a) and speed--density (b) planes for 
data in the left lane together with the line $J$ (a) and the curve $J$ (b) associated with
the propagation of the downstream front of the wide moving jam. (c--g) --
Traffic phases in the flow--density (c, e, g) and speed--density (d, f, h) planes for 
{\it improved data} in the left (c, d), middle (e, f), and right (g, h) lanes.
Moving averaging over the platoon of five vehicles is used.
\label{Em_Average}  }
\end{center}
\end{figure*}

A comparison of average characteristics of low vehicle speeds within a wide moving jam
associated with moving blanks (crossing and triangle points) with free flow (black quadrates), synchronized flow
(circles), and the line $J$ for the downstream jam front (line $J$)
is presented in Fig.~\ref{Em_Average}, in which
a moving averaging of vehicle platoons of five   vehicles passing the detector has been performed.
As   found earlier~\cite{Neubert,Knospe2002,Knospe2004}
(see Fig. 1 in~\cite{Knospe2004}), very small density states associated with low vehicle speeds  are observed within 
wide moving jams (crossing points in Fig.~\ref{Em_Average} (a, b)). Due to these $\lq\lq$characteristic" small density states within wide moving jams
(crossing points in Fig.~\ref{Em_Average} (a, b))
the flow rate is usually in average an increasing function of  density
in the flow--density plane (crossings plus triangle points in Fig.~\ref{Em_Average} (a)).

However, a  single vehicle 
 analysis of these small density points that has been made allows us to conclude that these points  
(crossing points in Fig.~\ref{Em_Average} (a)) as well as
the related increasing function of the flow rate with density associated with
  wide moving jams~\cite{Knospe2004}    result from a {\it large systematic error}.
  This error is
associated with    error density estimation within the jams made in~\cite{Neubert,Knospe2002,Knospe2004}.
To understand this critical conclusion,
recall that due to traffic flow interruption within the wide moving jam
some of the   time headways are very large (Fig.~\ref{WideJam} (b, d, f)).
The use of 
the single vehicle data associated with  large time headways for 
the density   estimation
  through the formula $\rho=q/v$ ($q$ and $v$ are the flow rate and   average speed)
  leads to very small densities with the jam. 
  However, during traffic flow interruption within the jam vehicle do not move;
    the real density of these standing vehicles is very large.
  This is because the related measured   time headways are not related to   time headways
  between  vehicles {\it moving} in   traffic. Thus, the single vehicle data associated with 
  the traffic flow interruption effect within wide moving jams 
  cannot be used for   density estimation within the jams.
  To estimate
  real vehicle density within a wide moving jam during  flow interruption,
the  density definition (number of vehicles per a freeway length at a given   time moment), i.e.,
{\it spatial averaging}
  should be used, rather than density estimation through the use of the formula  $\rho=q/v$ in which
  the flow rate and average speed are related to {\it time averaging} of vehicles passing a detector
  during a given   time interval. The spatial averaging (density definition) is not possible to apply, when
  data is measured at detectors whose locations are far away from
   each other (Fig.~\ref{Single_Fig} (a)).
  The conclusion about the large systematic error in density estimation within wide moving jams
  is confirmed by a numerical analysis made in Sect.~\ref{S_B_Sect}.
  
 This analysis shows also that within wide moving jams  density estimation $\rho=q/v$ 
 exhibits a relative small error during non-interrupted vehicle motion associated with moving blanks
 within the jams. This is because in this case
 the associated  time headways are   related to    vehicles  moving  within the jams.
  If only these single vehicle data are used for
density estimation, then   rather than an increasing function of the flow rate with density 
  (crossing and triangle points in Fig.~\ref{Em_Average} (a)), the  points for moving blanks within the jam
   associated with random transformations in different directions
  in the flow--density plane appear
  (triangle points in Fig.~\ref{Em_Average} (c, e, g)).
   This is associated with the physical meaning of
   low vehicle speed states  whose occurrence is due non-regular vehicle motion covering
   blanks within the wide moving jam.
   
   In Fig.~\ref{Em_Average}, points for synchronized flow are related to synchronized flow states
   in the jam inflow. The speed in these states is considerably higher than the speed within
   low speed states associated with moving blanks within the jam.
   For this reason, three types of points (for free flow, synchronized flow, and moving blanks within
   the wide moving jams) appear that are separated one from another in the flow--density and speed--density planes.
   These separated traffic states for the three traffic phases -- free flow, synchronized flow,
    and  wide moving jams -- are in accordance with the states on 
     a theoretical double Z-characteristics of
    three-phase traffic theory (see Sect. 6.4 in~\cite{KernerBook}), if
    low speed states within the wide moving jam are taken into account.
   
\begin{figure*}
\begin{center}
 \includegraphics[width=12 cm]{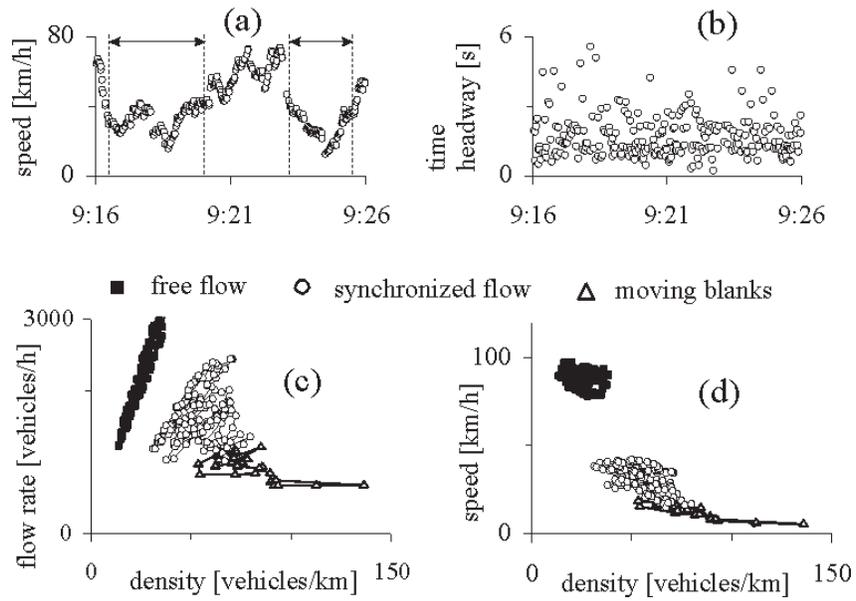}
\caption{Empirical characteristics of traffic phases: (a, b) -- 
Single vehicle data for speed (a) and time headways (b) for synchronized flow at D10 related to Fig.~\ref{Single_Fig} (b).
(c, d) -- Traffic states averaged over five vehicles (moving average) associated with synchronized flow in (a)
and with moving blanks within the jam taken from Fig.~\ref{Em_Average} (e, f)
in the flow--density (c) and speed--density planes (d). Arrows in (a) show time intervals of
synchronized flow states used in (c, d). Middle freeway lane.
\label{Em_Average_Syn}  }
\end{center}
\end{figure*}

   However,   synchronized flow states can overlap
   states associated with moving blanks within a wide moving jam (Fig.~\ref{Em_Average_Syn}).
   This is because the speed in synchronized flow states can be as low as the speed in states 
   associated with moving blanks
   within the jam.
   Thus, in general case low speed states associated with moving blanks within wide moving jams cannot be used for clear distinguishing 
   the synchronized flow and wide moving jam phases in congested traffic.

\subsection{Numerical Simulations of Moving Blanks}
\label{S_B_Sect}

In numerical simulations  presented below,
 we    use  the microscopic models   and model parameters
 of Ref.~\cite{KKl2003A,KKl2004AA}. The main idea of these models is the speed adaptation effect
 in synchronized flow that takes place when the vehicle cannot pass
 the preceding vehicle. Within a so-called $\lq\lq$synchronization gap",
 the vehicle tends to adjust its speed
 to preceding vehicle without caring, what the precise space gap is, as long as it is safe.
 Thus, there are an infinity of model steady states at each synchronized flow speed
 associated with the space gaps between the synchronization  and safe gaps, i.e.,
 there is no fundamental diagram for  steady state model solutions for synchronized flow in these models.
 In addition,  driver time delay depends on whether a vehicle 
 decelerates or accelerates as well as on the vehicle speed.
 This enable us to simulate  driver behavior in various traffic conditions.   
 A detailed consideration of the models, their physics and parameters can be found in Ref.~\cite{KKl2003A,KKl2004AA}
and
  Sects.   16.3 and 20.2 of the book~\cite{KernerBook}. When some other
 model parameters are   used, they are given in   figure captions.

\begin{figure*}
\begin{center}
\includegraphics[width=12 cm]{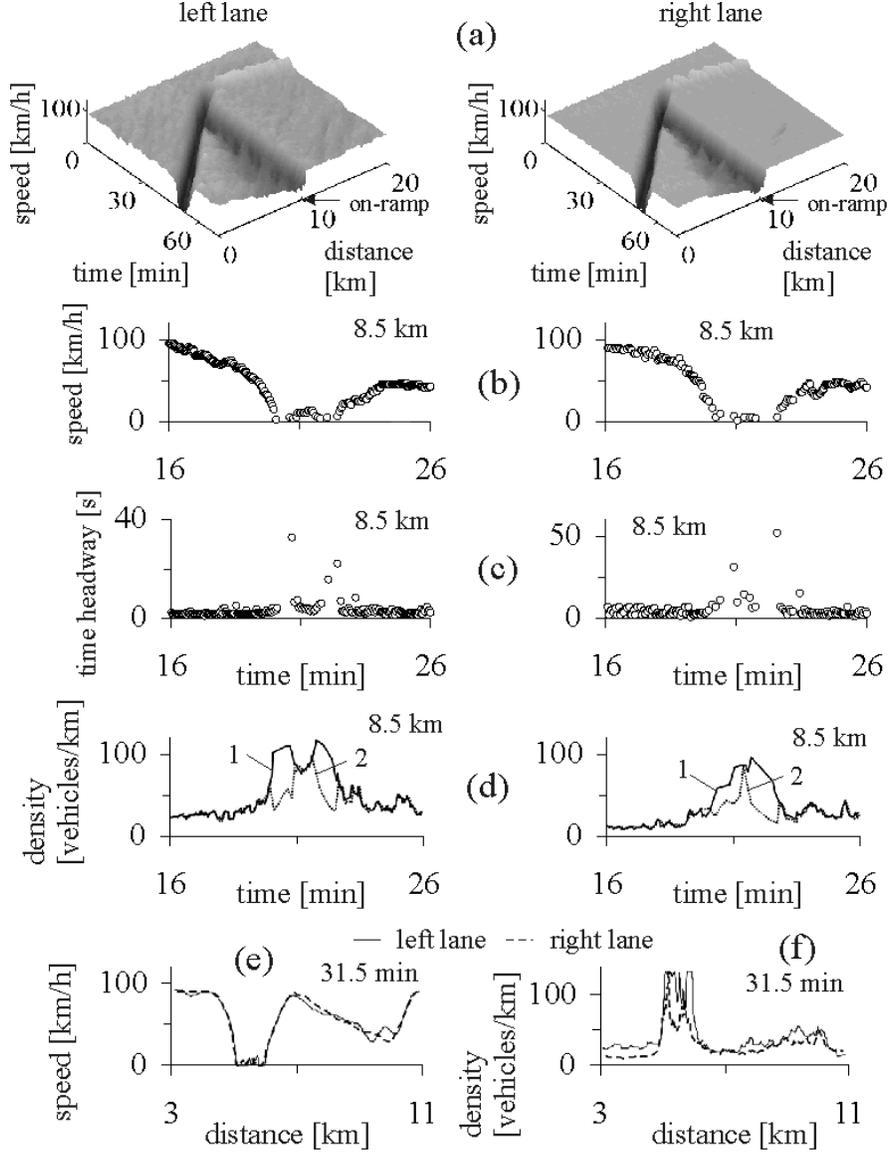}
\caption{Simulations of moving blanks within a wide moving jam propagating through
an on-ramp bottleneck. 
(a) -- Speed in space and time on the main road.
(b--d) -- Single vehicle data for time dependencies of speed (b), time headways (c),
 and the associated density distributions   calculated through the density definition
  (curves 1) and the formula $\rho=q/v$ (curves 2) (d) at the location 8.5 km.
Left and right figures (a--d) are related to the left and right lanes, respectively.
In (d) for density calculation,
  a moving averaging over the platoon of five vehicles
  is performed.
  To use the density definition by calculations of curves 1 in (d),
  at a time moment when one of the vehicles has just been detected at the location 8.5 km the density within a vehicle platoon
    is calculated. This platoon
     consists of the detected vehicle and two vehicles upstream and two vehicles
 downstream.
In contrast,  calculations of curves 2 in (d) 
are
performed as in empirical observations 
in Fig.~\ref{Em_Average}, i.e.,
   the flow rate $q$ and average speed $v$ of the platoon 
   passing the location are used in  the formula $\rho=q/v$.   (e, f) -- Speed (e) and density (f) related to (a) as functions of distance at the time moment
  31.5 min in the left and right lanes. 
  Model for heterogeneous traffic   of Sect.~20.2 in~\cite{KernerBook} with
80$\%$   fast  and 20$\%$   long vehicles;
$\tau^{\rm(a, \ j)}_{\rm del}(v)=\tau /p^{(j)}_{0}(v)$,
$p^{(j)}_{0}(v)=(a^{(j)} +b^{(j)}\min(1, \ v/10))$, $j=$ 1,3;
$a^{(1)}=0.565, \ b^{(1)}=0.085$ for fast vehicles ($j=$ 1),
and
$a^{(3)}=0.3, \ b^{(3)}=0.18$ for long vehicles ($j=$ 3).
$q_{\rm in}=$ 1565, $q_{\rm on}=$ 337 vehicles/h/lane.
$q_{\rm out}=$ 1900  and $q_{\rm out}=$ 1100
vehicles/h in the left and right lanes, respectively.
On-ramp location is $x_{\rm on}=$ 10 km,   on-ramp merging region length is
$L_{\rm m}=$ 300 m.
The wide moving jam is exited by  local perturbations applied in both lanes;
  perturbation duration $T^{(\rm pert)}$     and their
location  $\Delta x^{(\rm pert)}$ downstream of  the end of the on-ramp merging
region   are: (3 min, 1250 m).
\label{Mic_Cr_wide}  }
\end{center}
\end{figure*}

For a moving jam in Fig.~\ref{Mic_Cr_wide} (a, b),
the criterion (\ref{GrossDel}) is satisfied. 
 Indeed, there are two regions in which flow interruption within the jam occur.
 The maximum time headways (Fig.~\ref{Mic_Cr_wide} (c))
 associated with these flow interruption effects satisfy the criterion (\ref{GrossDel}) 
   (model time delay $\tau^{\rm (ac)}_{\rm del}\approx$ 1.74 sec)
 in the left and right freeway lanes. As a result, in accordance with the macroscopic spatiotemporal criteria for the 
  wide moving jam phase, this jam
propagates through an on-ramp bottleneck while maintaining
the mean velocity of the downstream jam front.

\begin{figure*}
\begin{center}
\includegraphics[width=12 cm]{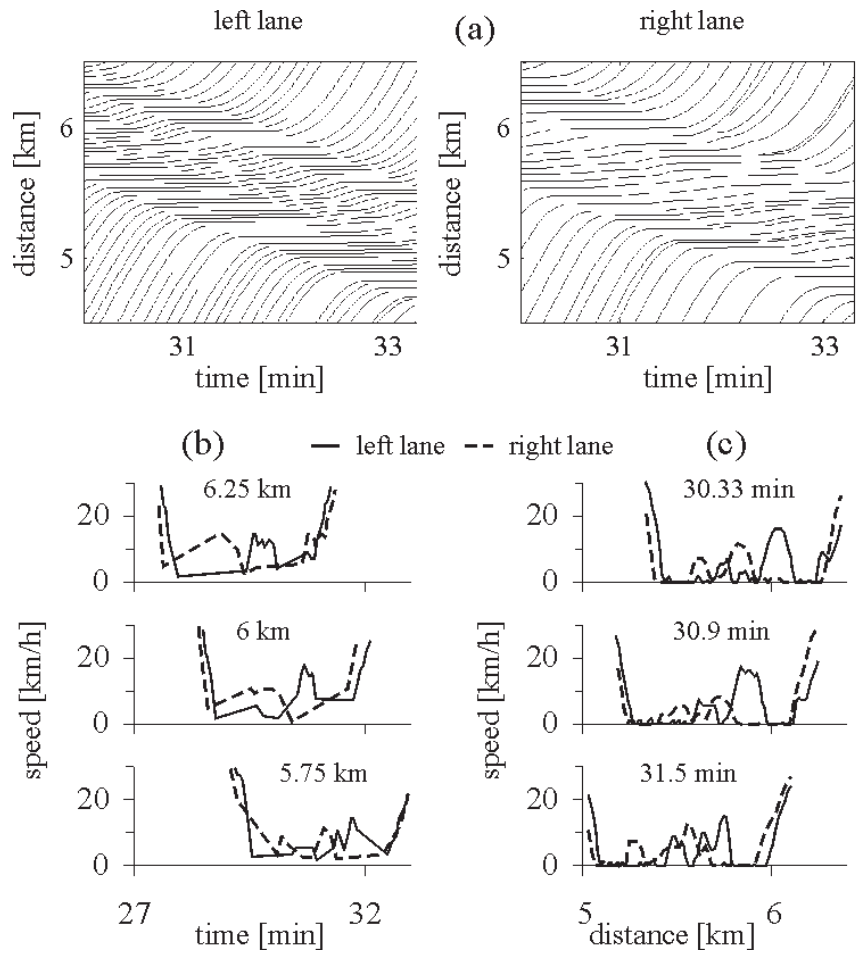}
\caption{Microscopic characteristics of moving blanks within the wide moving jam
shown in Fig.~\ref{Mic_Cr_wide} (a):
(a) -- Vehicle trajectories in the left (left) and right (right) lanes.
(b, c) -- Moving blanks at three different locations (b) and three different time moments (c)
in the left and right lanes. Trajectories of each 4th vehicle are shown.
 \label{BlanksWide} } 
\end{center}
\end{figure*}

 Within the wide moving jam, between time intervals of flow interruptions there are intervals
 in which low speed states associated with moving blanks (Fig.~\ref{Mic_Cr_wide} (b, c)).
 As can be seen from vehicle trajectories within the jam, moving blanks are related to a non-regular vehicle motion
 of covering blanks between vehicles within the jam (Fig.~\ref{BlanksWide} (a)).
 These moving blanks propagate upstream with a negative velocity that is in average equal to the
 characteristic speed of the downstream jam front. During jam propagation,
 locations of moving blanks as well as their spatial and temporal distributions exhibit complex variations within the jam,
 which are different in the left and right lanes
 (Fig.~\ref{BlanksWide} (b, c));
 these variations seem to correspond to a random vehicle speed behavior
 within the jam.
 
\begin{figure*}
\begin{center}
\includegraphics[width=12 cm]{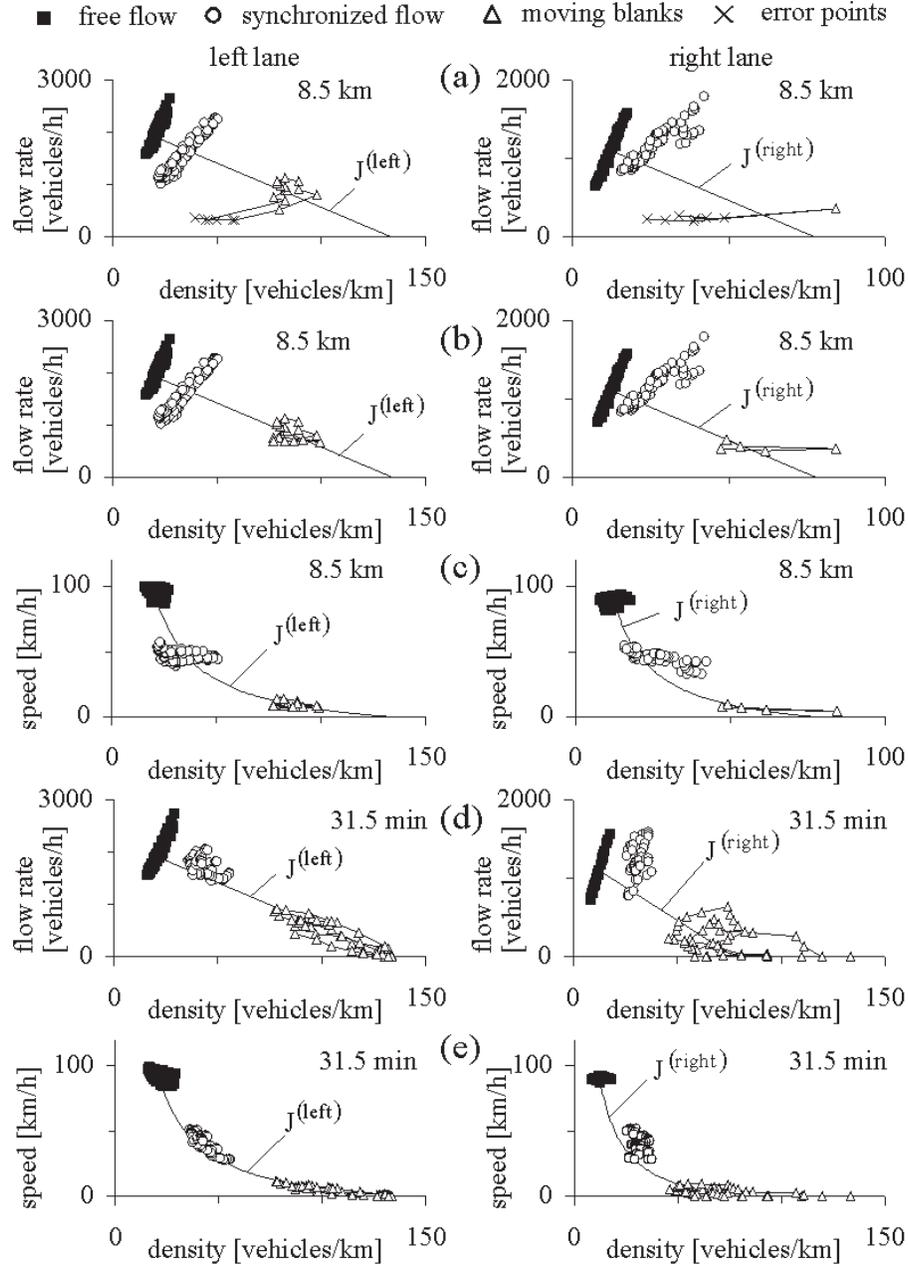}
\caption{Characteristics of moving blanks within the wide moving jam
shown in  Fig.~\ref{Mic_Cr_wide} (a) together with states
of free flow and synchronized flow in the flow--density
  (a, b, d) and speed--density planes (c, e). Left and right figures are related to the left and right lanes, respectively:
(a--c) -- States within the jam are determined at  detector 
located at 8.5 km through the formula $\rho=q/v$ in two cases in which
  all states within the jam measured at the detector  are shown (a) and error states  are removed (b, c).
(d, e) -- States within the jam are determined through the density definition (vehicles per freeway length) at  time moment
31.5 min. Moving average over five vehicles in vehicle platoons is used.
$J^{\rm (left)}$ and $J^{\rm (right)}$ are the line $J$   (a, b, d) and the associated curve $J$ (c, e)
for the downstream jam
front in the left and right lanes, respectively.
 \label{Mic_Average} } 
\end{center}
\end{figure*}

 Due to flow interruption within the wide moving jam shown in Fig.~\ref{Mic_Cr_wide} (a--c),
 there is a  large error in   density distributions calculated through the formula
 $\rho=q/v$ in the vicinity of the jam fronts at greater density (curves 2 in Fig.~\ref{Mic_Cr_wide} (d))
 in comparison with density distributions found based on
 the density definition (vehicles per freeway length) (curves 1). Indeed, curves 2 in Fig.~\ref{Mic_Cr_wide} (d)
 show a significant density underestimation within   moving jams.
 These error states (crossing points in Fig.~\ref{Mic_Average} (a)) explain the systematic error in the empirical studies
 of states within wide moving jams made in Ref.~\cite{Neubert,Knospe2002,Knospe2004} that has been illustrated in
 Fig.~\ref{Em_Average} (a, b) (compare error points in Figs.~\ref{Em_Average} (a, b) and~\ref{Mic_Average} (a)).
 
 Only within the regions of moving blanks  errors in
 density distributions calculated through the formula
 $\rho=q/v$   (curves 2 in Fig.~\ref{Mic_Cr_wide} (d))  
 in comparison with density distributions found based on
 the density definition   (curves 1) decrease considerably.
 For this reason, if error states associated with flow interruption within the jam are removed,
 then remaining states in the flow--density (Fig.~\ref{Mic_Average} (b))
 and speed--density planes  (Fig.~\ref{Mic_Average} (c)) exhibit
 a qualitative  correspondence with the states within the jam calculated through the density definition
 (Fig.~\ref{Mic_Average} (d, e)) 
and
related to the speed and density distributions
 found at a fixed time moment 31.5 min (Fig.~\ref{Mic_Cr_wide} (e, f)).
 Nevertheless,  remaining errors lead to some quantitative differences in 
 traffic state determinations within the jam based on detector measurements 
 (Fig.~\ref{Mic_Average} (b, c)) and on the density calculation through the density definition
 (Fig.~\ref{Mic_Average} (d, e)). In particular, there are points in the flow--density plane
 found in
 the   density calculation through the density definition (Fig.~\ref{Mic_Average} (d)), 
 which are related to  greater density up to the maximum jam density.
 These points cannot usually be found based on
 traffic measurements at a detector location (Fig.~\ref{Mic_Average} (b, c)).

\begin{figure*}
\begin{center}
\includegraphics[width=12 cm]{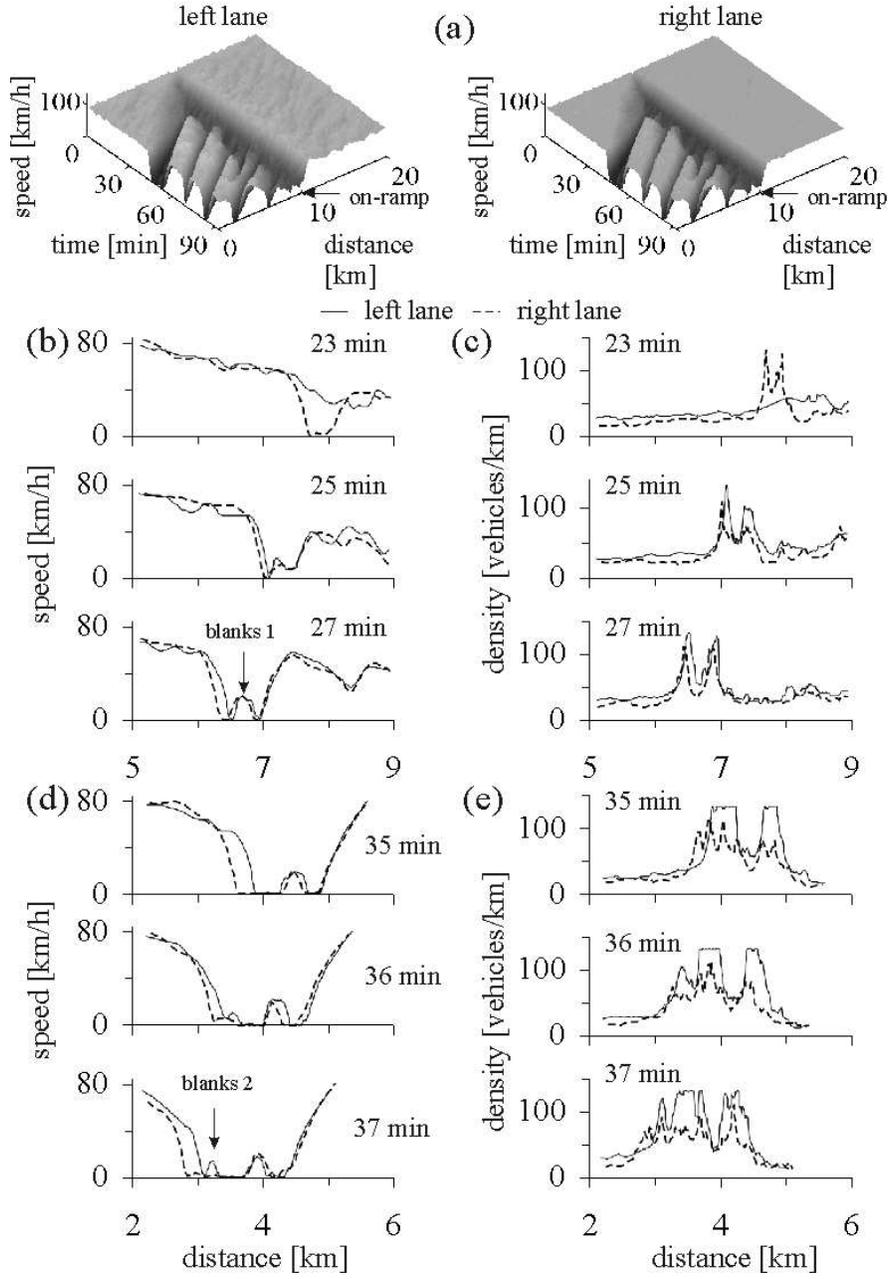}
\caption{Simulations of moving blanks emergence: (a) --
Speed in space and time within general pattern (GP) at on-ramp bottleneck in the left (left) and right lanes (right).
(b--e) -- Speed (b, d) and density (c, e) determined through the use of the density definition (vehicles per freeway length)
at different time moments associated with the GP in (a) in the left and right lanes.
$q_{\rm in}=$ 1714, $q_{\rm on}=$ 450 vehicles/h/lane.
Model and bottleneck parameters  are the same as those in Fig.~\ref{Mic_Cr_wide}.
 \label{Blanks_jam_in_GP} } 
\end{center}
\end{figure*}

\begin{figure*}
\begin{center}
\includegraphics[width=12 cm]{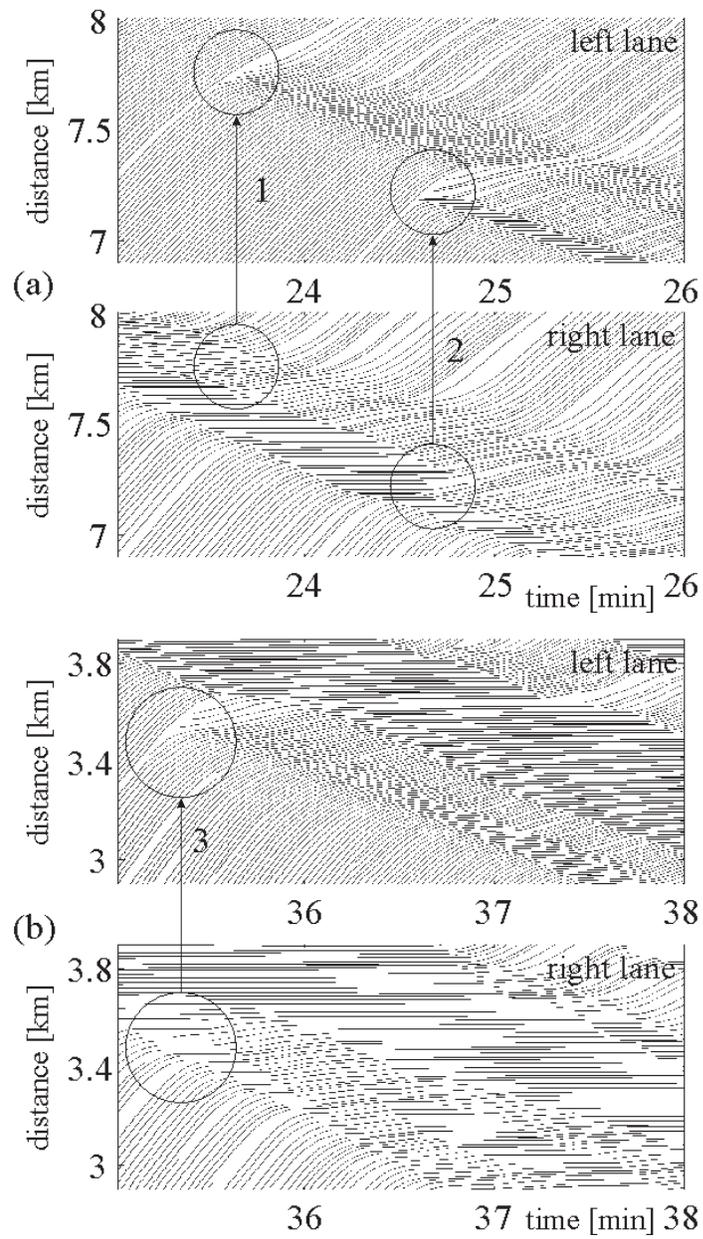}
\caption{Moving blanks emergence in the GP shown in Fig.~\ref{Blanks_jam_in_GP} (a): (a, b) --
Vehicle trajectories for different time intervals 
 in the left   and right lanes.
 \label{Blanks_GP} } 
\end{center}
\end{figure*}

 To understand possible scenarios of moving blanks emergence within a wide moving jam,
 spontaneous jam emergence in synchronized flow of an GP at an on-ramp bottleneck
 has been simulated (Fig.~\ref{Blanks_jam_in_GP}).
 Due to a heterogeneous traffic consisting of   vehicles
 and long vehicles used in these simulations, a moving jam emerges firstly in the right lane only
 ($t=$ 23 min in Figs.~\ref{Blanks_jam_in_GP} (b, c) and~\ref{Blanks_GP} (a)).
 Then some vehicles in the right lane change to the left lane (arrow 1 in Fig.~\ref{Blanks_GP} (a)).
 One the one hand, this lane changing decreases speed in the left lane.
 On the other hand, space gaps (blanks) between vehicles 
in the right lane
increase due to lane changing.
 Vehicles in the right lane begin to cover these blanks 
 ($t\approx$ 24 min in Fig.~\ref{Blanks_GP} (a)).
 As a result, low speed states within the jam appear associated with these moving blanks.
 Later,  subsequent lane changing of vehicles from the right to the left lane results
 in an abrupt decrease in speed in the left lane leading to jam formation in this lane
 (arrow 2 in Fig.~\ref{Blanks_GP} (a)). This causes
 the associated abrupt jam front formation and moving blanks emergence 
 ($t\approx$ 25 min in Figs.~\ref{Blanks_jam_in_GP} (b, c) and~\ref{Blanks_GP} (a)).
 Thus, in this scenario vehicle lane changing from the right to the left lane
 at the upstream jam front is the main reason for moving blanks emergence within the jam
 (moving blanks labeled $\lq\lq$blanks 1" in Fig.~\ref{Blanks_jam_in_GP} (b)).
 
 During subsequent jam propagation new moving blanks  
 emerge  (Fig.~\ref{Blanks_jam_in_GP} (d, e)).
 Firstly, the upstream jam front in the right lane is upstream of the jam front in the left lane 
($t=$ 35 min, 
Fig.~\ref{Blanks_jam_in_GP} (d, e)). Due to lane changing of vehicles from the right to the left lane,
 the front locations are synchronized each other
 (arrow 3 in Fig.~\ref{Blanks_GP} (b)).
 This lane changing causes  moving blanks emergence within the jam
 ($t=$ 37 min, moving blanks labeled $\lq\lq$blanks 2" in Fig.~\ref{Blanks_jam_in_GP} (d)).

\section{Moving Jam Fronts and Jam Propagation Loops}
\label{Front_Sect}

\begin{figure*}
\begin{center}
\includegraphics[width=12 cm]{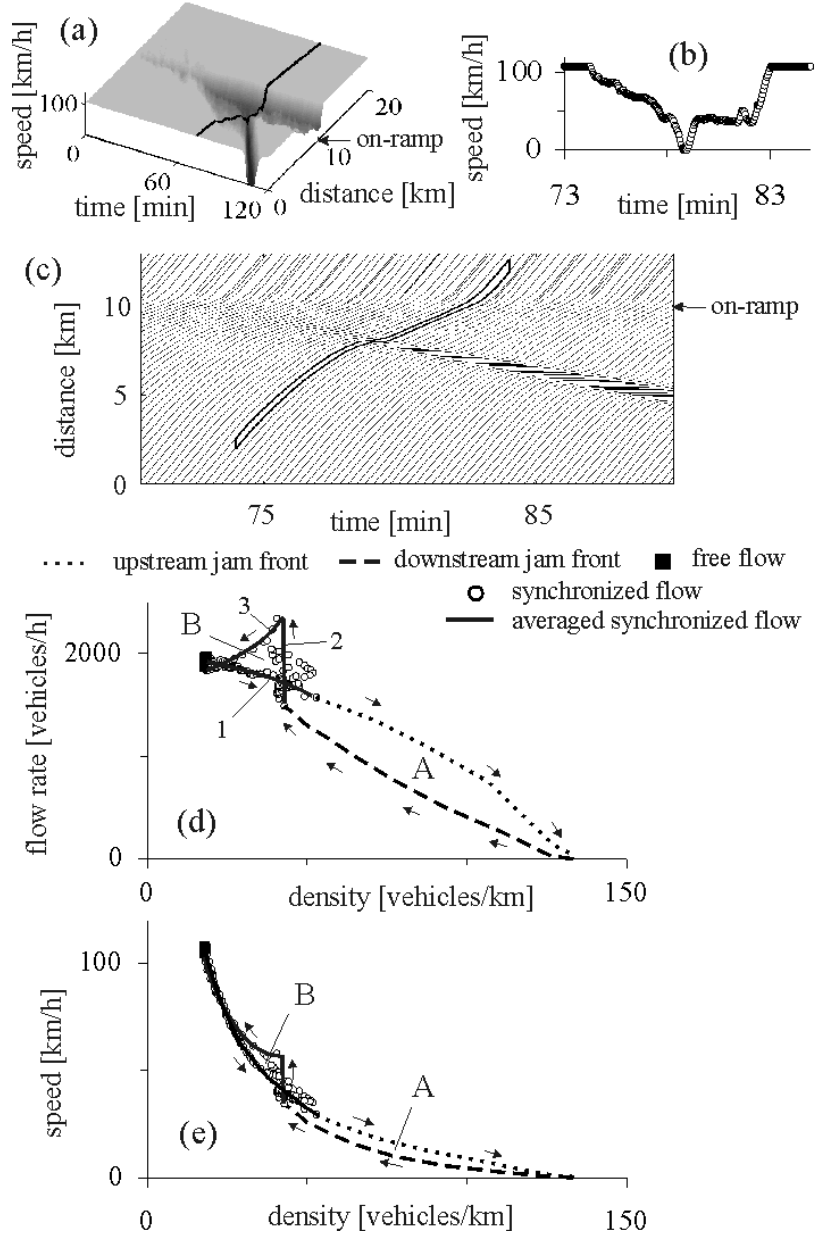}
\caption{Microscopic features of moving jam propagation:
(a) -- Speed on the main road in time and space related to a dissolving general pattern (DGP) that emerges 
spontaneously at an
on-ramp bottleneck.
(b) -- Speed within a platoon of seven vehicles
propagating through the wide moving jam and synchronized flow phases formed in the DGP; speed is averaged over the vehicle platoon.
Solid curve in (a) shows the average trajectory associated with   platoon propagation.
(c) -- Vehicle trajectories within the DGP in which the vehicle platoon related to (b) is marked by   solid block.
(d, e) -- Flow--density (d) and speed--density relationships (e) associated with the platoon of seven vehicles
going   through the wide moving jam and synchronized flow phases   related to the DGP in (a--c).
Single-lane model of identical vehicles (Sect. 16.3 of Ref.~\cite{KernerBook}).
$q_{\rm in}=$ 1946, $q_{\rm on}=$ 345 vehicles/h. $x_{\rm on}=$ 10 km,  $L_{\rm m}=$ 300 m.
In (c) trajectories of each 10th vehicle are shown.
\label{Mic_fronts_wide} } 
\end{center}
\end{figure*}

Whereas 
within the upstream jam front vehicles decelerate when they met the jam, within the downstream jam front
vehicles accelerate escaping from the jam. In simulations made on a single-lane road with an on-ramp bottleneck
(Fig.~\ref{Mic_fronts_wide}), a moving jam is
      formed in a dissolving general pattern (DGP) that emerges
    spontaneously  at the bottleneck. A platoon of seven vehicles
    is considered that is going through the   moving jam and other traffic states related to the DGP. Within the
      platoon the  density $\rho$ is calculated at each given time moment,
      i.e., in accordance with the density definition (vehicles per freeway length). 
      The flow rate associated with the platoon is 
      $q=\rho v$, where $v$ is average vehicle speed  within the platton. It turns out that  traffic states
      in the flow--density and speed--density planes associated with these speed and density within the platoon
      exhibit two loops labeled $A$ and $B$ in Fig.~\ref{Mic_fronts_wide} (d, e).    
      The loops are associated with different characteristics of the three
traffic phases, free flow, synchronized flow, and wide moving jam passing by the vehicle platoon
going through the DGP (solid curve in Fig.~\ref{Mic_fronts_wide} (a)).
       These numerical results (Fig.~\ref{Mic_fronts_wide} (d, e))
   explain empirical results of Treiterer
       and Myers~\cite{Treiterer1974A} and Treiterer~\cite{Treiterer1975} 
       found in their empirical study of vehicle trajectories within a moving
       jam, which emerged and propagated upstream of an on-ramp bottleneck~\cite{Hysteresis}.     

  To understand these statements, note that upstream of the bottleneck firstly synchronized flow occurs. Then
  a  narrow moving jam emerges spontaneously in this synchronized flow. During jam propagation,
  the jam amplitude grows and the narrow moving jam transforms into a wide moving jam, i.e.,
  an S$\rightarrow$J transition occurs in the synchronized flow. This wide moving 
  jam prevents subsequent moving jam emergence
  in synchronized flow 
  upstream of the bottleneck. As a result, the DGP appears at the bottleneck, which consists of
  synchronized flow 
upstream
of the moving jam, the moving jam, and synchronized flow that is
  downstream of the jam and upstream of the bottleneck (Fig.~\ref{Mic_fronts_wide} (a);
   see a more detailed consideration of the physics of DGP emergence in the book~\cite{KernerBook}).
  
  Thus, the chosen vehicle platoon (solid curve in Fig.~\ref{Mic_fronts_wide} (a))
  is going firstly through free flow ($t<74$ min
  in Fig.~\ref{Mic_fronts_wide} (a, b)). Secondly, the platoon must decelerate when it reaches
  synchronized flow, which is upstream of the wide moving jam. In this synchronized flow,
   speed decreases during the platoon propagation ($74<t<77$ min
  in Fig.~\ref{Mic_fronts_wide} (a, b)). The associated synchronized flow states 
  lie at lower speed and greater density   in the flow--density and speed--density planes 
  than the speed and density within the platoon
  in free flow, respectively (curve 1 in Fig.~\ref{Mic_fronts_wide} (d)).
  Thirdly, the platoon meets the upstream jam front within which the vehicles
  must decelerate up to the speed within the jam (this speed is almost as low as zero;
  $t\approx$ 78.5 min in Fig.~\ref{Mic_fronts_wide} (a, b)). Consequently,
  the flow rate and speed sharply decrease and   density increases     
   (curve for the upstream jam front in Fig.~\ref{Mic_fronts_wide} (d, e)). Later, due to upstream jam propagation through the platoon,
   the vehicles within the platoon can accelerate
      at the downstream jam front   ($t\approx$ 79 min in Fig.~\ref{Mic_fronts_wide} (a, b)).
      As a result,  
  the flow rate and speed sharply increase and   density decreases     
   (curve for the downstream jam front in Fig.~\ref{Mic_fronts_wide} (d, e)). We see that at the upstream jam front
   the flow rate is greater than at the downstream front at the same density. This is because
      vehicles decelerating at the upstream jam front
     accept shorter time headways than  time headways of   vehicles accelerating
    at the downstream jam front.
    
  Then the platoon moves through synchronized flow  downstream of the jam. In this synchronized flow,
   the speed is lower than in free flow
  upstream of the jam and downstream of the bottleneck ($79<t<82$ min in Fig.~\ref{Mic_fronts_wide} (a, b)). 
  This platoon propagation   
  is also associated  with the synchronized flow within the merging region of the on-ramp bottleneck
  (Fig.~\ref{Mic_fronts_wide} (a--c)). Within this region due to the vehicle merging from the on-ramp onto the main road
  time headways within the platoon  decrease. Therefore, the
  flow rate   is a sharply increasing   function of  freeway location.
  For this reason, traffic states in the flow--density and speed--density planes
   associated with the platoon, which is going through synchronized flow downstream of the jam,
  exhibit rapid flow rate growth with a slightly increasing speed.  
  These synchronized flow states 
  lie at lower speed and greater density  in the flow--density and speed--density planes 
  than the speed and density in free flow, respectively (curve 2 in Fig.~\ref{Mic_fronts_wide} (d)).
  This growth of the flow rate   results in the loop labeled $A$
  in the flow--density and speed--density planes.
  This loop
   is formed by  the upstream and downstream jam fronts together with a part of the curve 2
  related to the synchronized flow downstream of the jam (Fig.~\ref{Mic_fronts_wide} (d, e)).
  
  Later, the platoon propagates through the downstream front
   of synchronized flow in which the speed increases considerably up to the speed in free flow downstream of the bottleneck
   ($t\approx$ 82.5 min in Fig.~\ref{Mic_fronts_wide} (b)).
   Hence, traffic states in the flow--density and speed--density planes
   related to the platoon propagating through the downstream front of synchronized flow at the bottleneck 
  exhibit rapid speed growth    in the flow--density and speed--density planes 
   (curve 3 in Fig.~\ref{Mic_fronts_wide} (d)). 
   Consequently, the platoon propagation through synchronized flow upstream of the jam (curve 1), 
   through synchronized flow
   downstream of 
   the jam (curve 2), as well as through the downstream front of the synchronized flow at the bottleneck up to free flow downstream of this front
   (curve 3) lead to the second loop labeled $B$ in the flow--density and speed--density planes 
   (Fig.~\ref{Mic_fronts_wide} (d)).

\begin{figure*}
\begin{center}
\includegraphics[width=12 cm]{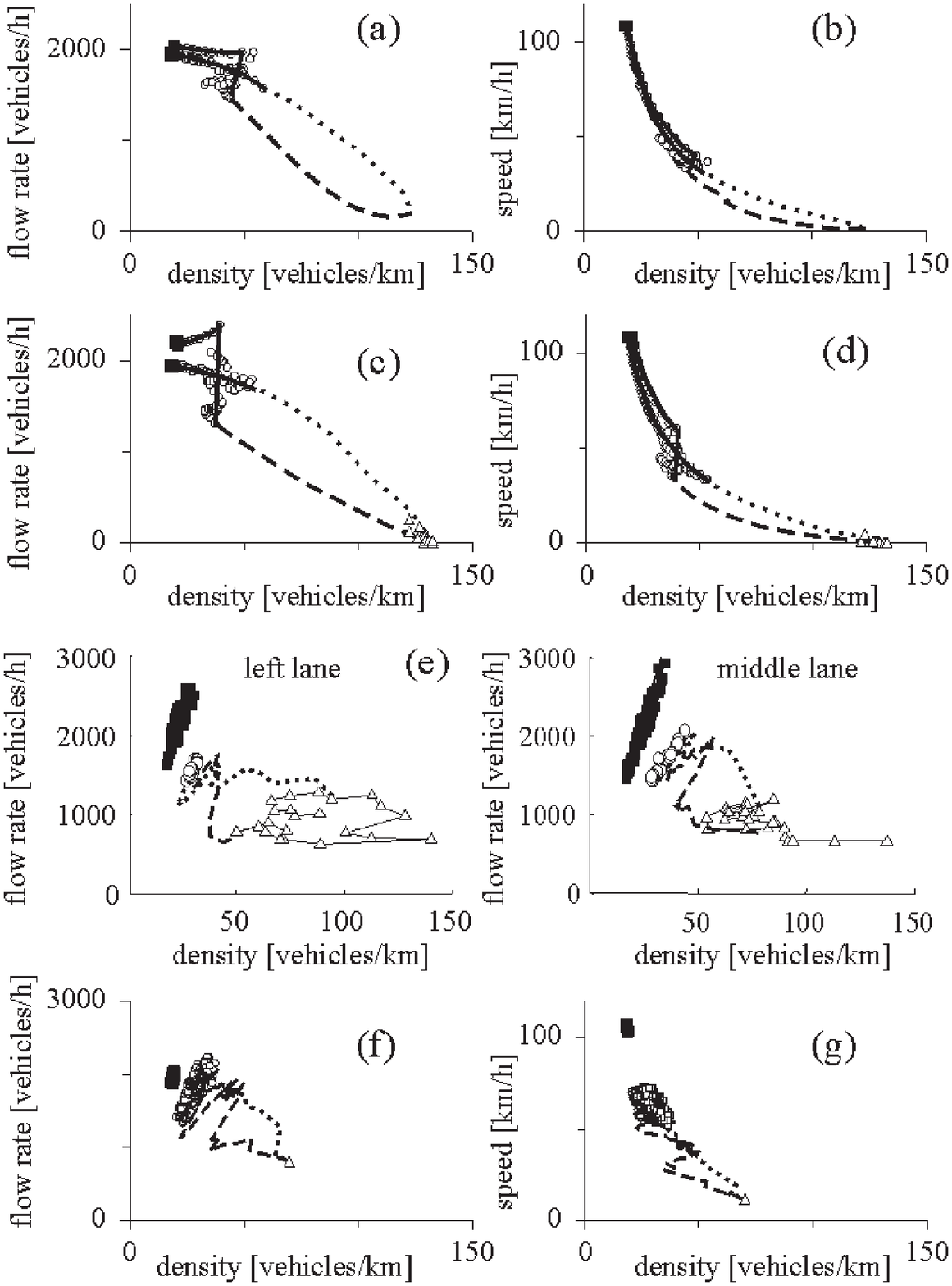}
\caption{Simulated flow--density (a, c) and speed--density relationships (b, d) of two different platoons of seven vehicles
going through the wide moving jam and other traffic phases within the DGP shown in Fig.~\ref{Mic_fronts_wide} (a). 
(e) -- Empirical flow--density  relationships in the left and middle lanes associated with the wide moving jam in Fig.~\ref{WideJam}.
(f, g) -- Simulated flow--density (f) and speed--density relationships (g)  associated with  moving averaging
of speed $v$ and flow rate $q$ within platoons of seven vehicles going through a virtual detector at $x=$ 8 km within the DGP
shown in Fig.~\ref{Mic_fronts_wide} (a). 
\label{Mic_fronts_wide_detectors} } 
\end{center}
\end{figure*}

       Simulations show that   loops' shapes depend on  a vehicle platoon that is chosen.
       If a platoon of   seven vehicles is chosen, which propagates earlier through the DGP
       than the platoon considered in Fig.~\ref{Mic_fronts_wide} (b--e),   then
         the loops $A$ and $B$ (Fig.~\ref{Mic_fronts_wide_detectors} (a, b))
         are different from those in   Fig.~\ref{Mic_fronts_wide} (d, e). There are several reasons for this result. Firstly,
   free flow   and synchronized flow states  
    upstream and downstream of the jam within the platoon depend on the chosen platoon considerably.
   Secondly, in Fig.~\ref{Mic_fronts_wide_detectors} (a, b)
         the platoon is going through  is a narrow moving jam rather than through a wide moving one.

      If another platoon of   seven vehicles  is chosen, which propagates later through the DGP
       than the platoon considered in Fig.~\ref{Mic_fronts_wide} (b--e), then
       the shape of the loops $A$ and $B$ (Fig.~\ref{Mic_fronts_wide_detectors} (c, d) is  different
      from those in   Figs.~\ref{Mic_fronts_wide} (d, e) 
      and~\ref{Mic_fronts_wide_detectors} (a, b). The first reason for this different behavior is the same as those mentioned above.
      In addition,
      in this case the platoon is going through   a wide moving jam whose width is greater
      than the width of the wide moving jam associated with the platoon chosen in Fig.~\ref{Mic_fronts_wide} (b--e).
      As a result, there are low speed states within the jam associated with moving blanks within the jam.
     
     In general, due to upstream jam propagation the distance between the downstream jam front  and  bottleneck location
       is a time function; in addition, jam characteristics change over time considerably. Thus,
       average space gaps and speeds within a platoon depend on the time interval, when the platoon is going
       through a moving jam, through synchronized flows 
      and free flows upstream and downstream of the jam.   This 
      can change both  loops $A$ and $B$ in  Fig.~\ref{Mic_fronts_wide_detectors} 
(a--d)
      associated with  a vehicle platoon going through even the same DGP
       appreciably. Moreover,
         the flow rate in   free flow downstream of the bottleneck
   associated with a vehicle platoon can be very different for different platoons. This is due to considerable variations in
    space gaps between vehicles within different platoons after the platoons have passed
   the downstream front of synchronized flow   at the bottleneck. This explains
   different flow rates in the free flow associated with three vehicles platoons in 
   Figs.~\ref{Mic_fronts_wide} 
(d)
and~\ref{Mic_fronts_wide_detectors}
(a, c).  
   Apparently,
       this complex loop behavior explains very different loops $A$ and $B$   observed in the empirical study
        by Treiterer
       and Myers~\cite{Treiterer1974A} and Treiterer~\cite{Treiterer1975}.

       The loops  in the flow--density and speed--density planes
       shown in Fig.~\ref{Mic_fronts_wide} (c, d) and~\ref{Mic_fronts_wide_detectors}
 (a--d)
       are associated with measurements of the density within the vehicle platoon in accordance with
       the density definition (vehicles per freeway length). The use of the density
       definition is possible, if vehicle trajectories  are known as this the case in simulations presented in the article above
       or in the Treiterer
       and Myers's empirical observations
        of vehicle trajectories~\cite{Treiterer1974A,Treiterer1975}. 
        
        Unfortunately, the use of this correct procedure
        is not possible to perform, when single vehicle data  
         is measured by a local detector
       on a road as this the case for the wide moving jam shown in Fig.~\ref{WideJam}.
        In this case, the density has to be estimated through the formula $\rho=q/v$, in which
        the flow rate $q$ and speed $v$ are associated with time averaging over many {\it different}
        vehicle platoons of seven vehicles passing the detector.
        There are two systematic errors in this case: (i) Density estimation at lower speeds within the jam
        (see explanations of Fig.~\ref{Mic_Cr_wide} (d) in Sect.~\ref{S_B_Sect}).
        (ii) Measured and estimated   traffic variables 
        associated with
        different platoons passing the detector do not coincide with the respective    traffic variables  found for 
  a single vehicle platoon going through the wide moving jam and other traffic phases in Fig.~\ref{Mic_fronts_wide}. 
  
  Apparently these errors lead to  qualitative different empirical
    loops  in the flow--density and speed--density planes (Fig.~\ref{Mic_fronts_wide_detectors} (e)) for the jam  shown in Fig.~\ref{WideJam}
    in comparison with empirical observations by Treiterer
       and Myers~\cite{Treiterer1974A,Treiterer1975}. These conclusions
       are confirmed
       by numerical simulations in which the same procedure of the density estimation through detector
       measurements is used (Fig.~\ref{Mic_fronts_wide_detectors} (f, g))
       as those in empirical Fig.~\ref{Mic_fronts_wide_detectors} (e).
       The simulated loops  in the flow--density and speed--density planes (Fig.~\ref{Mic_fronts_wide_detectors} (f, g)) 
       associated with traffic variables determined at a  detector location within the DGP shown in
    Fig.~\ref{Mic_fronts_wide} (a) are qualitatively different from
    the loops  found through the use of 
    the correct density calculation associated with vehicle platoon propagation through the DGP
     (Fig.~\ref{Mic_fronts_wide} (d, e)).

\section{Discussion
\label{Discussion}}

Based on empirical and model results presented, the following conclusions can be made:

(i) In empirical observations and numerical simulations,
moving blanks that are responsible for low speed states within a wide moving jam
exhibit complex a non-regular spatiotemporal   behavior. The blanks are often not synchronized between different
freeway lanes.

(ii) 
Lane changing at the upstream front of a wide moving jam
can lead to large moving blanks within the jam.

(iii) Moving blanks emergence is characterized by  
random changes of the upstream jam front location in some of freeway lanes
over time.

(iv) To find   correct characteristics associated with 
moving blanks within   wide moving jams, measurement points related to 
the flow interruption effect within the jams leading to a large 
systematic error in density estimation  have to be removed, when
wide moving jam characteristics are calculated.

(v) In simulations, there are two loops in the flow--density
plane associated with a vehicle platoon propagation through
a moving jam as well as through
free flow and synchronized flow states of traffic upstream and downstream of the jam.
The loops are associated with different characteristics of the three
traffic phases, free flow, synchronized flow, and wide moving jam passing by the platoon.
These simulation results made in the context of three-phase traffic theory explain
empirical observations of traffic dynamics in vehicle trajectories  
found by   Treiterer
       and Myers~\cite{Treiterer1974A} and Treiterer~\cite{Treiterer1975}.

\begin{thebibliography}{8.}
\bibitem{Aggregated}

Sometimes for better statistical conclusions,
traffic variables obtained
from various days or/and many different locations are aggregated to a data set; this data base of aggregated data is further used for 
 calculation of  traffic flow characteristics.
\bibitem{MovingJams}

From a comparison of
a local traffic dynamics measured at 
D7 with the dynamics at D10
in Fig.~\ref{Single_Fig} (b)
it has been found that the moving jams propagate upstream with the mean velocity of the downstream
jam fronts $v_{\rm g}\approx -$ 15.8 km/h.
\end{thebibliography}
\end{document}